\documentclass[fleqn,twoside]{article}
\usepackage{espcrc2}
\usepackage{epsfig}

\usepackage{graphicx}
\usepackage[figuresright]{rotating}

\newcommand{\AmS}{{\protect\the\textfont2
  A\kern-.1667em\lower.5ex\hbox{M}\kern-.125emS}}

\title{
{\normalsize\hfill{\vspace*{-1.5cm}
         \vbox{\hbox{CERN-TH/2002-295}
         \hbox{YITP-SB-02-52}
         \hbox{IFIC/02-49}
          \hbox{hep-ph/0210359}}}}\\
\vskip 1.5cm
Theory of Neutrino Masses and Mixing
\thanks{
Plenary talk given at the 31st International Conference on 
High Energy Physics, ICHEP02 (Amsterdam,24-31 July, 2002).} 
}
\author{M. C. Gonzalez-Garcia \address
  {Theory Division, CERN, CH-1211, Geneva 23, Switzerland,\\
  Y.I.T.P., SUNY at Stony Brook, Stony Brook,NY 11794-3840\\
and IFIC, Universitat de Val\`encia -- C.S.I.C., Apt 22085, 46071
  Val\`encia, Spain}}
\begin{document}
\begin{abstract}
In this talk I will review our present knowledge on neutrino masses
and mixing trying to emphasize  
what has been definitively proved and what is in 
the process of being probed. I will also discuss the most
important theoretical implications of these results: 
the existence of new  physics, the estimate of the scale 
of this new physics as well as some other possible consequences
such as leptogenesis origin of the baryon asymmetry.
\vspace{1pc}
\end{abstract}

\maketitle

\section{Introduction}
\begin{figure}
\begin{center}
\includegraphics[scale=0.5]{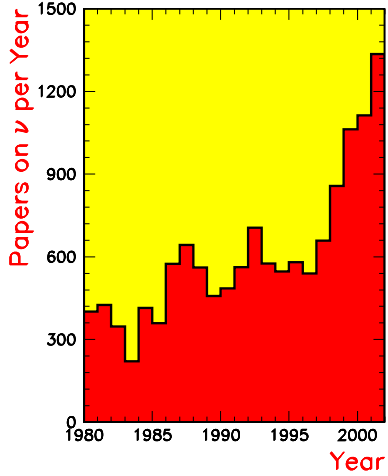}
\end{center}
\vglue -1.5cm
\caption{Number of papers in SPIRES with the word ``neutrino'' 
in the title as a function of the year.}
\label{papers}
\end{figure}

Neutrino physics is a very exciting field at this moment. From the
plenary talk by D. Wark~\cite{wark} as well as from the 
talks by S. Oser~\cite{sno}, M.R. Vagins~\cite{sksolar}, 
T. Mitsui~\cite{kland}, C. Mauger~\cite{skatm}, Y. Hayato~\cite{k2k},
Y. Itow~\cite{jhf}, A. Bazarko~\cite{miniboone} and J. Urheim~\cite{minos}
in the neutrino parallel session we have heard about the enormous
progress made and being made in the experimental front from which
we have learned that 
\begin{itemize}
\item  Solar $\nu_e's$ convert to $\nu_{\mu}$ or
$\nu_\tau$. This evidence was first established at 3.4~$\sigma$ by the 
comparison of the SNO results~\cite{sno} on the charged current (CC)  
measurement with the results from SuperKamiokande(SK) on the electron 
scattering (ES)~\cite{sksolar} of $^8$B neutrinos, 
and  with more than 5$\sigma$ 
from the subsequent SNO neutral current (NC)  and CC observations~\cite{sno}. 
These results are new since the last ICHEP conference in Osaka.
\item The evidence of atmospheric $\nu_\mu$  disappearing from SK is now at
$> 15 \sigma$, most likely converting to $\nu_\tau$~\cite{skatm}.
K2K data~\cite{k2k} supports within statistics the disappearance of
$\nu_\mu$'s. The most likely explanation is neutrino oscillations. 
\item  LSND found evidence for 
$\overline{\nu_\mu}\rightarrow\overline{\nu_e}$ which is being tested
by MiniBooNE~\cite{miniboone}. 
\end{itemize}
The experimental results have triggered a very intense 
activity in the phenomenological and theoretical front. 
In Fig.~\ref{papers}  I show 
the number of papers in
SPIRES with the word ``neutrino'' in the title as a
function of the year where one can see the clear {\sl forward peak}
corresponding to the last two years since the last ICHEP conference. 
This figure makes it clear that it is impossible to summarize in full
fairness the activity of the field. 
I will concentrate on reviewing the phenomenological interpretations 
and some of the theoretical implications of the experimental results. 

The outline of the talk is as follows: 
in Sec.~\ref{osci} I will  briefly review the notation and parameter
space of neutrino oscillations. In Sec.~\ref{fits}
I will discuss the phenomenological interpretation of solar, atmospheric and
laboratory experiments in terms of neutrino masses and mixing  
emphasizing the still-existing ambiguities
and uncertainties in the interpretation. Sec.~\ref{impli}
is devoted to some of  the theoretical implications.  
Finally in Sec.~\ref{future} I will 
discuss what we will learn in the near future from existing experiments
and what will be still left to learn at proposed facilities.   
\subsection{Leptonic Mixing and $\nu$ Oscillations}
\label{osci}
If neutrinos have masses, flavour is mixed in the  CC interactions
of the leptons, and a leptonic mixing matrix will
appear analogous to the CKM~\cite{ckm} matrix for the quarks.
The possibility of arbitrary mixing between two massive neutrino 
states was first introduced in Ref.~\cite{MNS}.
The discussion of leptonic mixing in generic models is complicated
by two factors. First the number 
massive neutrinos ($n$) is unknown, since there are
no constraints on the number of right-handed, SM-singlet,
neutrinos ($m=n-3$). Second, since neutrinos carry neither 
color nor electromagnetic charge, they could be Majorana
fermions. Denoting the neutrino mass eigenstates by $\nu_i$,
$i=1,2,\ldots,n$, and the charged lepton mass eigenstates by $(e,\mu,\tau)$,
in the mass basis, leptonic CC interactions are given by
\begin{equation}
-{\cal L}_{\rm CC}={g\over\sqrt{2}}(\overline{e_L}\ \overline{\mu_L}\ 
\overline{\tau_L})\gamma^\mu U\pmatrix{\nu_1\cr 
.\cr .\cr \nu_n\cr} W_\mu^+ +{\rm h.c.}.
\label{CClepmas}  
\end{equation} 
Here $U$ is a $3\times n$ matrix. 

Given the charged lepton mass matrix $M_\ell$ and the neutrino mass matrix
$M_\nu$ in some interaction basis [the 
interaction eigenstates are denoted by  $(e^I,\mu^I,\tau^I)$ and 
$\vec\nu=(\nu_{Le},\nu_{L\mu},\nu_{L\tau},\nu_{s1},\dots,\nu_{sm})$]
\begin{equation}
-{\cal L}_{M}=(\overline{e_L^I}\ \overline{\mu_L^I}\ 
\overline{\tau_L^I})\ M_\ell \pmatrix{e_R^I\cr\mu_R^I\cr\tau_R^I\cr}
+ \frac{1}{2}\overline{\vec\nu^c} M_\nu \vec\nu +{\rm h.c.}\; ,
\end{equation}
we can find the diagonalizing matrices $V^\ell$ ($3\times 3$) and 
$V^\nu$ ($n\times n$):
\begin{eqnarray}
{V^\ell}^\dagger M_\ell M_\ell^\dagger V^\ell=
{\rm diag}(m_e^2,m_\mu^2,m_\tau^2),\ \ \  \nonumber \\
{V^\nu}^\dagger M_\nu^\dagger M_\nu V^\nu={\rm diag}(m_1^2,m_2^2,m_3^2,\dots,
m_n^2) 
\label{findia}  
\end{eqnarray} 
The $3\times n$ mixing matrix $U$ can be found from these 
diagonalizing matrices:
\begin{equation}
U_{ij}=P_{\ell,ii}\, {V^\ell_{ik}}^\dagger \, V^\nu_{kj}\, (P_{\nu,jj}).
\label{diamat}  
\end{equation}
$P_\ell$ is a diagonal $3\times3$ phase matrix, that is conventionally
used to reduce by three the number of phases in $U$.
$P_\nu$ is a diagonal matrix with additional arbitrary phases (chosen
to reduce the number of phases in $U$) only for Dirac states. 
For Majorana neutrinos, this matrix is simply a unit matrix. 
The reason for that is that if one rotates a Majorana neutrino 
by a phase, this phase will appear in its
mass term which will no longer be real.
Thus,  the number of phases that can be absorbed by redefining the
mass eigenstates depends on whether the neutrinos are Dirac or Majorana 
particles. In particular, if there are only three Majorana (Dirac) neutrinos, 
$U$ is a $3\times 3$ matrix analogous to the CKM matrix for the
quarks
but due to the Majorana (Dirac) nature of the neutrinos it depends on 
six (four) independent parameters: three mixing angles and three (one) phases.

In absence of new interactions for the charged leptons, 
we can identify their interaction eigenstates with 
the corresponding mass eigenstates after phase redefinitions, 
and the CC lepton mixing matrix 
$U$ is simply given by a $3\times n$ sub-matrix of the unitary 
matrix $V^\nu$.  

The presence of the leptonic mixing, allows for flavour oscillations
of the neutrinos~\cite{pontecorvo}. 
A neutrino of energy $E$ produced in a 
CC interaction with a charged lepton $l_\alpha$ can be detected 
via a CC interaction with a charged lepton $l_\beta$ with a probability
\begin{eqnarray}
{ P_{\alpha\beta}} &=&\delta_{\alpha\beta}-4\sum_{i< j}^n
{ \mbox{Re}[U_{\alpha i}U^*_{\beta i} U^*_{\alpha j} U_{\beta j}]}
\sin^2x_{ij}
\nonumber \\
&&+2
\sum_{i<j}^n
{\mbox{Im}[U_{\alpha i}U^*_{\beta i} U^*_{\alpha j} U_{\beta j}]}
\sin^2\frac{x_{ij}}{2}
 \; ,  \label{pab}
\end{eqnarray}
where 
$x_{ij}=1.27 \frac{\Delta m^2_{ij}}{\rm eV^2} \frac{L/E}{\rm m/{\rm MeV}}\;$ ,
with  $\Delta m^2_{ij} \equiv m_i^2-m_j^2$.
$L=t$ is the distance between the production point of
$\nu_\alpha$ and the detection point of $\nu_\beta$.
The first line in  Eq.~(\ref{pab}) is CP conserving while the second one
is CP violating and has opposite sign for $\nu$ and $\bar\nu$.

The transition probability  [Eq.~(\ref{pab})] presents an oscillatory 
behaviour, with oscillation lengths  
$L_{0,ij}^{\rm osc}=\frac{4 \pi E}{\Delta m_{ij}^2}$
and amplitude that is proportional to elements in the mixing matrix. 
From Eq.~(\ref{pab}) we find that neutrino oscillations are only sensitive
to mass squared differences. Also, the Majorana phases cancel out and only the
Dirac phase is observable in the CP violating term. 
In order to be sensitive to a given value of 
$\Delta m^2_{ij}$, an experiment has to be set up with $E/L\approx 
\Delta m^2_{ij}$ ($L\sim L_{0,ij}^{\rm osc}$). 

For a two-neutrino case, the mixing matrix depends on a single parameter,
there is a single mass-squared difference $\Delta m^2$ and there is 
no Dirac CP phase.
Then $P_{\alpha\beta}$ of Eq.~(\ref{pab}) takes the well known form  
\begin{equation}
P_{\alpha\beta}=\delta_{\alpha\beta}- (2\delta_{\alpha\beta}-1) \sin^22\theta 
\sin^2x \;.
\label{ptwo}
\end{equation} 
The full physical parameter space is covered with $\Delta m^2\geq 0$ 
and $0\leq\theta\leq\frac{\pi}{2}$ (or, alternatively,
$0\leq\theta\leq\frac{\pi}{4}$ and either sign for $\Delta m^2$).
Changing the sign of the mass difference, $\Delta m^2\to-\Delta m^2$, and
changing the octant of the mixing angle, $\theta\to\frac{\pi}{2}-\theta$,
amounts to redefining the mass eigenstates, $\nu_1\leftrightarrow\nu_2$:
$P_{\alpha\beta}$ must be invariant under such transformation. 
Eq.~({\ref{ptwo}) reveals, however, that $P_{\alpha\beta}$ is actually
invariant under each of these transformations separately. This situation 
implies that there is a two-fold discrete ambiguity in the interpretation
of $P_{\alpha\beta}$ in terms of two-neutrino mixing: the two different sets of
physical parameters, ($\Delta m^2, \theta$) and ($\Delta m^2, \frac{\pi}{2}
-\theta$), give the same transition probability in vacuum. One cannot tell from
a measurement of, say, $P_{e\mu}$ in vacuum whether the larger component of 
$\nu_e$ resides in the heavier or in the lighter neutrino mass eigenstate.

This symmetry is lost when neutrinos travel through regions of dense matter. 
In this case, they can undergo forward scattering with
the particles in the medium. These interactions are, in general, 
flavour dependent and they can be included as a potential
term in the evolution equation of the flavour states. As a consequence 
the oscillation pattern is modified. Let us consider, for instance, 
oscillations $\nu_e\rightarrow \nu_\mu$ in a neutral medium
like the Sun or the Earth. For this system, the instantaneous
mixing angle in matter takes the form
\begin{equation} 
{\sin 2\theta_{m}}= 
\frac{{ \Delta{m}^2 \sin 2\theta }} 
{\sqrt{ ({\Delta{m}^2 \cos 2\theta } -A)^2 
+({ \Delta{m}^2 \sin 2\theta} )^2 }} 
\label{effmix}
\end{equation}
where $A=2 E V_{\rm CC}=2 \sqrt{2} E G_F N_e$ ($N_e$ is the electron number density
in the medium). Eq~(\ref{effmix}) shows an enhancement (reduction) 
of the mixing angle in matter for $\theta<\frac{\pi}{4}$ 
($\theta>\frac{\pi}{4}$)~\cite{msw}. Thus, matter effects allow to 
determine whether the larger component of 
$\nu_e$ resides in the lighter neutrino mass eigenstate.
As we will see this is the presently favoured scenario 
for solar neutrino oscillations. 
For mixing of three or more neutrinos, the oscillation probability,
even in vacuum, does not depend in general of $\sin^2 2\theta_{ij}$.

Neutrino oscillation experiments measure $P_{\alpha\beta}$. It is common
practice to interpret these results in the two-neutrino
framework and translate the  constraints on $P_{\alpha\beta}$ 
into allowed or excluded regions in the plane 
($\Delta m^2,\; \sin^22\theta$). 
However, as we have seen once matter effects are important, or mixing among 
more than two neutrinos is considered, the covering of the full parameter
space requires the use of a single-valued function of the mixing angle
such as $\sin^2\theta$ or $\tan^2\theta$~\cite{dark}. 
\section{Global Fits}
\label{fits}
\subsection{Solar Neutrinos}
The sun is a source of $\nu_e's$ which are produced in the different
nuclear reactions taking place in its interior. Along this talk I will 
use the $\nu_e$ fluxes from Bahcall--Pinsonneault 
calculations~\cite{bp00} which I refer to as the solar standard model (SSM).
These neutrinos have been detected at the Earth by seven
experiments which use different detection techniques~\cite{wark}.
Due to the different energy threshold and the different detection
reactions, the experiments  are sensitive to different parts of the 
solar neutrino spectrum and to the flavour composition of the beam.
In table~\ref{tab:solarexp} I show the different experiments and detection
reactions with their energy threshold  as well as
their latest results on the total event rates as compared to 
the SSM prediction. 
\begin{table*}
\begin{center}
\begin{tabular}
{llccc}
 Experiment & { Detection}  & { Flavour} 
& { $ E_{\rm th}$ (MeV)} & { $\frac{\rm Data}{\rm BP00}$} \\[+0.1cm]
\hline 
{ Homestake} &  { $^{37}$Cl$(\nu,e^-)^{37}$Ar} & { $\nu_e$} 
& { $E_\nu> 0.81$} & { $0.34\pm 0.03$}\\[+0.1cm]  
{ Sage +}  & & &  &  \\ [-0.2cm]
& { $^{71}$Ga$(\nu,e^-)^{71}$Ge} & { $\nu_e$}
& { $E_\nu> 0.23$} & { $0.56\pm 0.04$}\\ [-0.2cm] 
{ Gallex+GNO} & & &  &  \\[+0.2cm]
&   &      { $\nu_e$, $\nu_{\mu/\tau}$} & &  \\[-0.2cm]
{ Kam     
$\Rightarrow$ SK} &  
{ ES $\;\nu_x e^- \rightarrow \nu_x e^-$} &  & 
{ $E_e > 5$} & { $0.46\pm 0.02$} \\[-0.2cm]
&            & {\footnotesize
$\left(\frac{\sigma_{\mu\tau}}{\sigma_e}\simeq \frac{1}{6}\right)$} 
&            &        \\[+0.1cm]
{ SNO} & { CC $\;\nu_e d \rightarrow p p e^-$} &
${\nu_e}$ &{ $T_e>5$}  & { $0.35\pm 0.02$} \\
  & { NC $\;\nu_x d \rightarrow \nu_x d$} &
{ $\nu_e$, $\nu_{\mu/\tau}$} 
&{ $T_\gamma >5$}  & { $1.01\pm 0.12$} \\
  & { ES $\;\nu_x e^- \rightarrow \nu_x e^-$} &
{ $\nu_e$, $\nu_{\mu/\tau}$} 
 &{ $T_e >5$}  & { $0.47\pm 0.05$} \\
\end{tabular}
\end{center}
\label{tab:solarexp}
\caption{Event rates observed at solar neutrino experiments compared to the
SSM predictions (the errors do not include the theoretical uncertainties).
For SNO, the quoted rates are obtained under the  hypothesis of 
undistorted $^8$B spectrum.}
\end{table*}
We can make the following statements: 
\begin{itemize} 
\item Before the NC measurement at SNO all experiments observed a flux 
that was smaller than the SSM predictions, $\Phi^{\rm obs}/\Phi^{\rm
  SSM}\sim0.3-0.6$. 
\item The deficit is not the same for the various experiments, 
which indicates that the effect is energy dependent.
\item SNO has observed an event rate different in the different reactions.
In particular in NC SNO observed no deficit as compared to the SSM.
\end{itemize}
The first two statements constitute the solar neutrino problem. The last one,
has provided us in the last year with evidence of 
flavour conversion of solar neutrinos independent of the solar model.

Both SK and SNO measure the high energy $^8$B neutrinos. 
Schematically, in presence of flavour conversion the observed fluxes
in the different reactions are
\begin{eqnarray}
\Phi^{\rm CC}&=&\Phi_e, \nonumber \\
\Phi^{\rm ES}&=&\Phi_e\,+\, r\,\Phi_{\mu\tau}, \\
\Phi^{\rm NC}&=&\Phi_e+ \Phi_{\mu\tau}, \nonumber 
\end{eqnarray}
where 
$r\equiv \sigma_{\mu}/\sigma_{e}\simeq 0.15$ 
is the ratio of the the $\nu_e - e$ and $\nu_{\mu} - e$ elastic 
scattering cross-sections. The flux
$\Phi_{\mu\tau}$ of active no-electron neutrinos is zero in the SSM.
Thus, in the absence of flavour conversion, the three observed 
rates should be equal.  The first reported SNO CC result 
compared with the ES rate from SK showed that the hypothesis of
no flavour conversion was excluded at $\sim 3\sigma$. 
Finally, with the NC measurement at SNO one finds that 
\begin{equation}
\Phi_{\mu\tau}=
(3.41\pm 0.45 ^{+0.48}_{-0.45}) \times 10^6\ {\rm cm}^{-2} {\rm s}^{-1}.
\label{snosk}
\end{equation}
This result provides evidence for neutrino flavor transition (from
$\nu_e$ to $\nu_{\mu,\tau}$) at the level of $5.3\sigma$. This
evidence is independent of the solar model.

The most generic and popular explanation to this observation is 
in terms of neutrino masses and mixing leading to oscillations 
of $\nu_e$ into an active ($\nu_\mu$ and/or  $\nu_\tau$) or a sterile 
($\nu_s$) neutrino.  Several global analyses of the solar neutrino data
have appeared in the literature after the latest SNO results~\cite{solana}.
In Fig.~\ref{solarosc} I show the results of a global 
analysis~\cite{oursolar} of the latest solar neutrino data in 
terms of oscillation parameters. 

To illustrate the progress in the last two years, I show in the same figure
the allowed parameter space which I showed in my talk at the ICHEP00 
conference two years ago~\cite{ichep00}. The progress can be summarized 
as follows:
\begin{itemize}
\item
Two years ago for the case of active--active neutrino oscillations we found three 
allowed regions for the global fit: the SMA solution, the LMA and 
LOW-QVO solution. The best solution was LMA but the other solutions were there
at 95\% CL. At 99\% CL the LMA region extended beyond maximal mixing 
and to $\Delta m^2$ above $10^{-3}$ eV$^2$. 
For sterile neutrinos the fit was slightly worse (due to the lack of neutral current
contribution to SK) but still reasonable.
\item
At present we find that active oscillations are clearly favoured. LMA is the
best fit and the only solution at $\sim$ 99\%CL. At 3$\sigma$ the allowed
parameter  space  within LMA is in the first octant  
and there is an upper bound  $\Delta m^2\leq 4\times 10^{-4}$ eV$^2$.
SMA is ruled out at $\sim 4\sigma$ as a consequence of the tension between
the low rate observed by SNO in CC and the flat spectrum observed by SK.  
Sterile oscillations are disfavoured at $\sim 5 \sigma$ 
due to the difference between the observed CC and NC event rates at
SNO.
\end{itemize}
In making this progress both the more detailed information on the
day-night spectrum of SK and the new SNO results have played very 
important and complementary roles.
\begin{figure}[ht]
\mbox{\epsfig{file=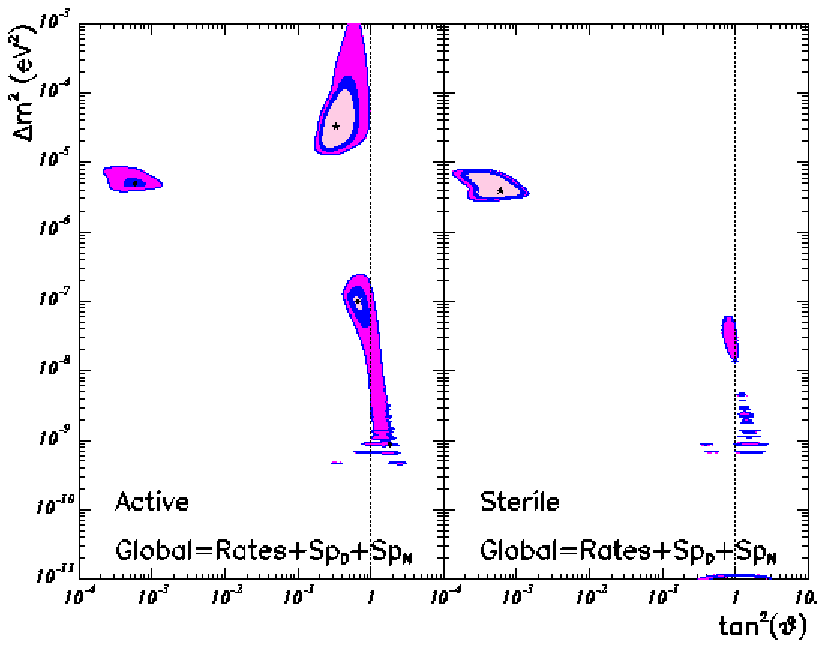,width=0.45\textwidth}} 
\mbox{\epsfig{file=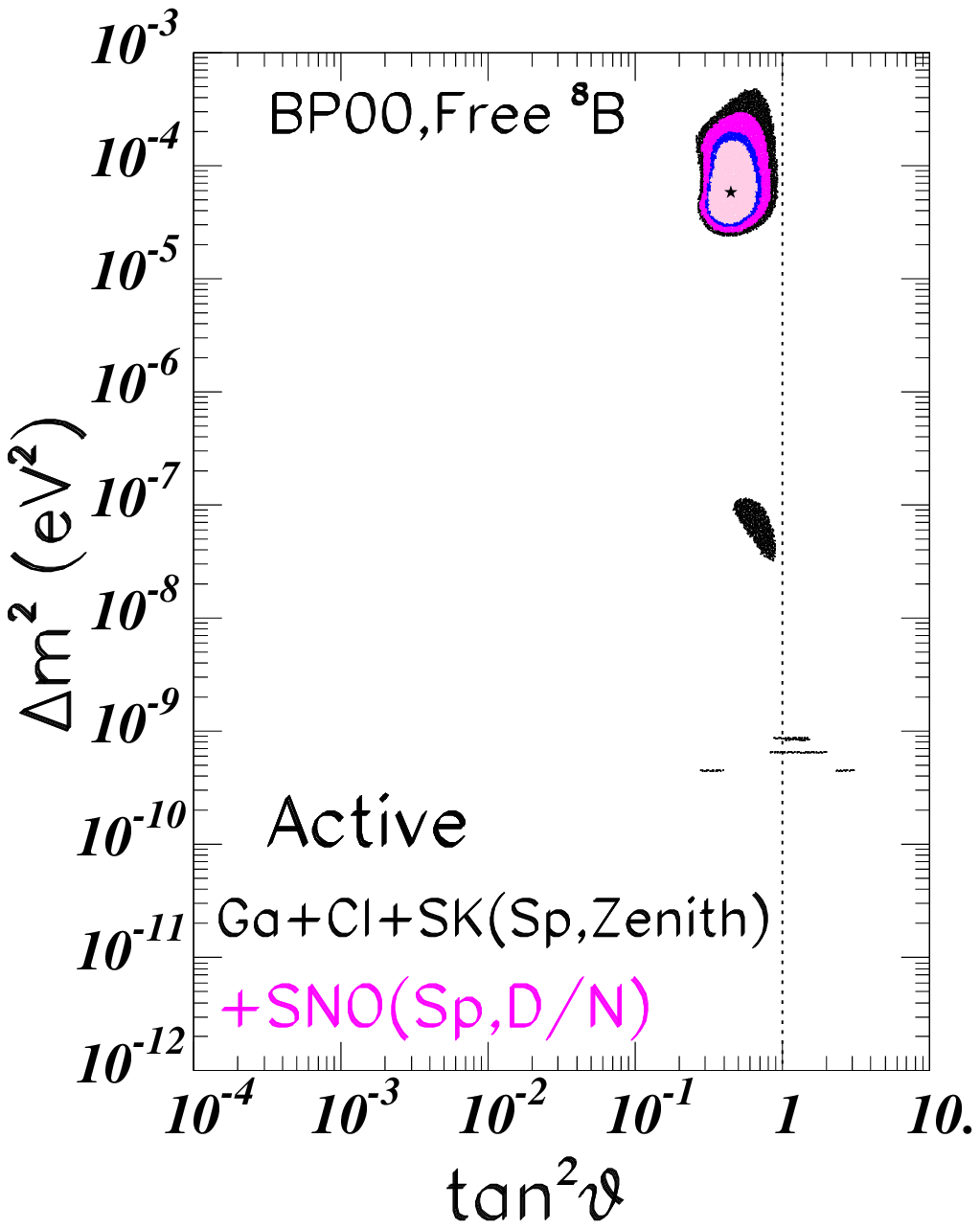,width=0.3\textwidth}} 
\mbox{\epsfig{file=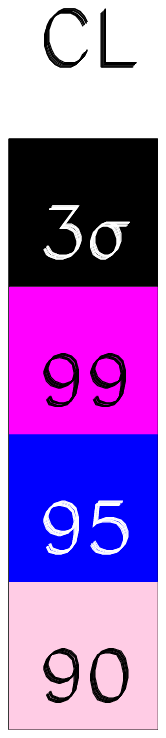,width=0.1\textwidth}} 
\vglue -1.cm
\caption{Allowed regions from the global fit for solar neutrino oscillations.
The upper panels show the status at the time of the ICHEP00 conference and 
the lower panel gives the results from the last updated analysis 
in Ref.~\cite{oursolar}.}
\label{solarosc}
\end{figure}

Next we can ask ourselves how certain we are that the flavour conversion
mechanism for solar neutrinos is indeed oscillations. 
To answer this question we can study which information we have on the 
actual energy dependence of the flavour conversion probability from 
the solar neutrino data in a model independent way
~\cite{solarprobs}. To do so one can fit the observed rates assuming 
different average survival probabilities
in three different energy ranges of the solar neutrinos: low energy
neutrinos, whose largest flux is the  $pp$ flux, 
with survival  probability $\langle P_{ee}\rangle_L$, 
a region of intermediate energy, 
consisting of the $^7$B, $pep$ and  CNO  fluxes, with survival probability 
$\langle P_{ee}\rangle_I$, and the higher energy part, 
whose dominant
contribution is due to $^8$B neutrinos, 
with survival probability $\langle P_{ee}\rangle_H$.  
I show the results of this exercise in Fig.~\ref{sunprob} together with the
predicted energy dependence of the survival probability at the best fit point
of LMA and LOW oscillations as well as for other alternative scenarios of
solar neutrino flavour conversion. 
\begin{figure}
\begin{center}
\includegraphics[scale=0.4]{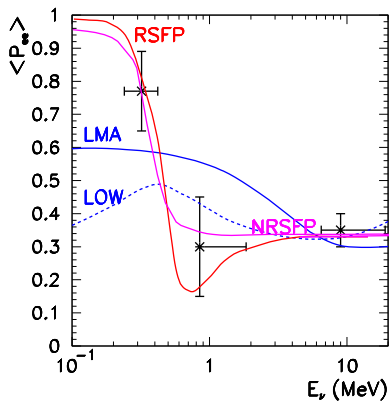}
\end{center}
\vglue -1.cm
\caption{Reconstructed values of the 
survival probability of solar neutrinos in different energy ranges 
from a fit to the observed
rates, together with the predictions from different flavour conversion mechanisms.} 
\label{sunprob}
\end{figure}

While the LMA oscillation may provide the simplest explanation of the 
data, there are  presently alternative scenarios which fit the data 
equally well.
For example spin-flavor precession~\cite{sfp} 
in which the neutrinos have an anomalous transition magnetic moment, which  
allows them to interact coherently with the magnetic field in the  
Sun. This can lead to resonant
as well as non-resonant  (RSFP and NRSFP respectively in 
Fig.~\ref{sunprob}) 
flavour transitions of neutrinos into anti-neutrinos
which, as seen in the figure can describe the observations 
(for appropriate choice of the magnetic field configuration),    
but implying masses and  mixing different than oscillations~\cite{sfpana}. 
For instance, for the case of RSFP the analysis of the data  gives 
values of $\Delta m^2=(0.8 - 1.5) \cdot 10^{-8} {\rm eV}^2$.

The mixing and level splitting required for neutrino flavour conversion 
can also be obtained in presence of non-standard neutrino interactions with 
matter~\cite{fcnc} which can lead to a correct description of the data 
even for massless neutrinos~\cite{fcncana}.

This alternative scenarios can be considered not very attractive from 
the theoretical point of
view, since they require some of the corresponding additional parameters (magnetic
moments and flavour changing couplings in the cases mentioned above) to take
``unnaturally'' (although still experimentally allowed) large values. However,
as I will discuss in 
Sec.~\ref{future} we are in the privileged situation that our theoretical
prejudices will soon become irrelevant since 
we have a running experiment, KamLAND
which, if it observes an oscillation signal, will rule out these scenarios
as the main mechanism of solar flavour conversion.
 
\subsection{Atmospheric Neutrinos}
Atmospheric showers are initiated when primary cosmic rays hit the
Earth's atmosphere. Secondary mesons produced in this collision,
mostly pions and kaons, decay and give rise to electron and muon
neutrino and anti-neutrinos fluxes whose interactions are 
detected in underground detectors~\cite{wark,skatm}.
\begin{figure}
\begin{center}
\includegraphics[scale=0.3]{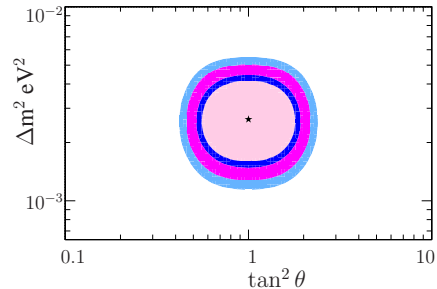}
\end{center}
\vglue -1cm
\caption{Allowed parameters from the global fit of atmospheric neutrino
  data for $\nu_\mu\rightarrow \nu_\tau$ oscillations 
from Ref.~\cite{michele}.}
\label{atmosc}
\end{figure}

At present the atmospheric neutrino anomaly (ANA) can be summarized in 
three observations: \\
-- There has been a long-standing deficit of about 60 \%  between the 
predicted and observed 
$\nu_\mu$/$\nu_e$ ratio of the contained events strengthened 
by the high statistics sample collected at the SK experiment. 
\\
-- The most important feature of the atmospheric neutrino
data at SK is that it exhibits a {\sl zenith-angle-dependent} deficit of 
muon neutrinos which indicates that the deficit is larger for muon neutrinos
coming from below the horizon which have traveled longer distances 
before reaching the detector. On the contrary, electron neutrinos behave
as expected in the SM.\\
-- The deficit for through-going muons is smaller that
for stopping muons, {\it i.e.} the deficit decreases as the neutrino 
energy grows.

The most likely solution of the ANA involves neutrino
oscillations. At present the best solution from a global analysis of the
atmospheric neutrino data is $\nu_\mu\rightarrow \nu_\tau$ 
oscillations with oscillation parameters shown in Fig.~\ref{atmosc} 

Oscillations into electron neutrinos are nowadays ruled out since
they cannot describe the measured angular dependence of muon-like
contained events. Moreover the most favoured range
of masses and mixings for this channel have been excluded by the 
negative results from the CHOOZ reactor experiment~\cite{chooz}.
Oscillations into sterile neutrinos are also disfavoured because
due to matter effects in the Earth they predict a flatter-than-observed  
angular dependence of the through-going muon data~\cite{skatm}.
\begin{figure}
\begin{center}
\includegraphics[scale=0.3]{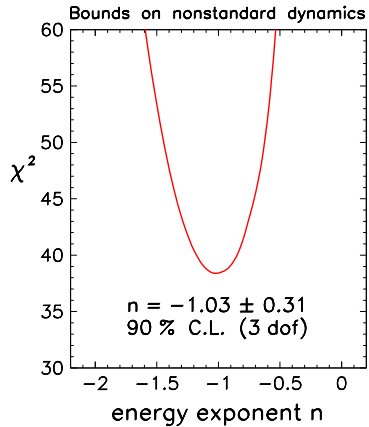}
\end{center}
\vglue -0.5cm
\caption{Results of a phenomenological fit of various flavour conversion  
mechanism to the  atmospheric neutrino data~Ref.~\cite{fogli}.}
\label{atmnoosc}
\end{figure}

Again, we can ask ourselves how certain we are that the flavour conversion
mechanism for atmospheric neutrinos is indeed oscillations.  
The answer is that the case for atmospheric neutrinos
oscillations can be stated with more confidence than the
case for solar neutrino oscillations, because the atmospheric
neutrino data spans over several decades in energy and distance which allows
a better discrimination between the oscillation hypothesis and other
flavour conversion mechanisms, which in general predict different dependence 
with energy and/or distance. In Fig.~\ref{atmnoosc} I show the results of 
the phenomenological analysis from the Bari Group~\cite{fogli}  
on which they fit the atmospheric data with a conversion probability 
$P_{\mu\tau}=\alpha \sin^2(\beta L E^n)$ which can parametrize several
conversion mechanisms. $n=-1$ corresponds to oscillations. Other mechanism
can lead to different values of the index $n$, for instance, violation of 
Lorenz invariance, or violation of the equivalence principle imply
$n=1$, n
on-universal neutrino coupling to a space-time torsion field
implies $n=0$.
The result of the fit shows that   $n=-1.03\pm 0.31$ at 90\%CL, clearly
favouring the hypothesis of oscillations as conversion mechanism.
\subsection{Three-Neutrino Oscillations}
So far I have discussed the evidence for neutrino
masses and mixings from solar and atmospheric data as usually formulated 
in the two--neutrino oscillation scenario. 
In Fig.~\ref{sum3fam} I show the summary of the allowed regions and channels
from solar and atmospheric data together with the 
bounds from the CHOOZ reactor experiment.
The minimum joint description of these data requires 
that all three known neutrinos take part in the oscillations. 
\begin{figure}
\begin{center}
\includegraphics[scale=0.35]{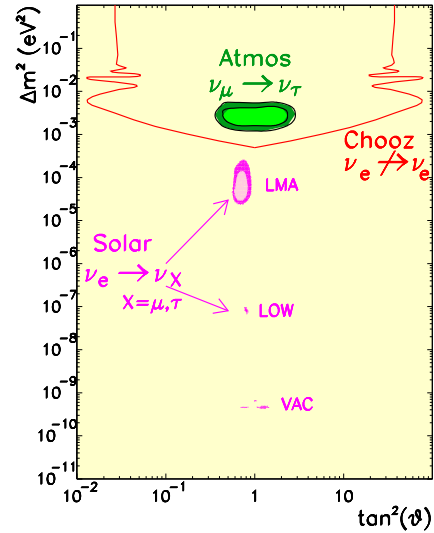}
\end{center}
\vglue -1cm
\caption{Pieces of the 3$\nu$ oscillation puzzle.}
\label{sum3fam}
\end{figure}

The mixing
parameters are encoded in the $3 \times 3$ lepton mixing matrix 
which can be conveniently parametrized 
in the standard form 
\begin{eqnarray}
U=\pmatrix{1&0&0 \cr 0& {c_{23}} & {s_{23}} \cr
0& -{s_{23}}& {c_{23}}\cr}\times
\nonumber  \\ \pmatrix{ 
{c_{13}} & 0 & {s_{13}}e^{i {\delta}}\cr
0&1&0\cr -{ s_{13}}e^{-i {\delta}} & 0  & 
{c_{13}}\cr}  
\times \pmatrix{{c_{21}} & {s_{12}}&0\cr
-{s_{12}}& {c_{12}}&0\cr
0&0&1\cr} \nonumber
\label{eq:evol.2} 
\end{eqnarray}
where $c_{ij}\equiv\cos\theta_{ij}$ and $s_{ij} \equiv \sin\theta_{ij}$.
Notice that, since the two Majorana phases do not affect neutrino 
oscillations, they are not included in the expression above.
The angles $\theta_{ij}$ 
can be taken without 
loss of generality to lie in the first quadrant, $\theta_{ij}\in[0,\pi/2]$.  

There are two possible mass orderings which, without any 
loss of generality can be chosen to be  as shown in Fig.~\ref{schemes}.
The direct scheme is naturally related to hierarchical masses,
$m_1\ll m_2\ll m_3$, for which $m_2\simeq\sqrt{\Delta m^2_{21}}$ and 
$m_3\simeq\sqrt{\Delta m^2_{32}}$, 
On the other hand, the inverted scheme implies that $m_3< m_1\simeq m_2$.
In both cases neutrinos can have  quasi-degenerate masses, $m_1\simeq 
m_2\simeq m_3\gg \Delta m^2_{21}, |\Delta m^2_{32}|$. 
The two orderings are often referred to in terms of the
sign($\Delta m^2_{31}$).   
\begin{figure}
\begin{center}
\includegraphics[scale=0.3]{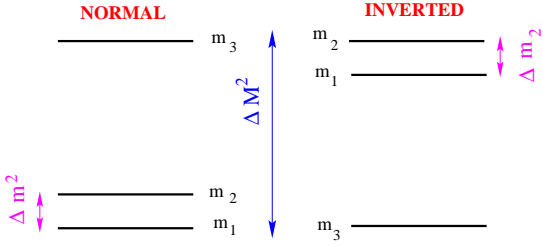}
\end{center}
\vglue -1cm
\caption{Mass schemes for 3 $\nu$ oscillations}
\label{schemes}
\end{figure}

In total the three-neutrino oscillation analysis involves 
seven parameters: 2 mass
differences, 3 mixing angles, the CP phase and the sign($\Delta m^2_{31}$).
Generic three-neutrino oscillation effects are:\\
-- Coupled oscillations with two different oscillation lengths, \\
-- CP violating effects, \\
-- Difference between Normal and Inverted schemes \\
The strength of these effects is controlled by the values of the ratio
of mass differences $\Delta m^2/\Delta M^2$, the mixing angle $\theta_{13}$ 
and the CP phase $\delta$. 

For solar and atmospheric oscillations, 
the required mass differences satisfy
\begin{equation}
\Delta m^2_\odot=\Delta m^2 \ll \Delta M^2=\Delta m^2_{\rm atm}.
\label{deltahier}
\end{equation}
Under this condition, the joint three-neutrino analysis simplifies and\\
-- For solar neutrinos the oscillations with the atmospheric oscillation 
length are averaged out and the survival probability takes the form:
\begin{equation}
P^{3\nu}_{ee,MSW}
=\sin^4\theta_{13}+ \cos^4\theta_{13}P^{2\nu}_{ee,MSW} 
\label{p3}
\end{equation}
where $P^{2\nu}_{ee,MSW}$ 
is obtained with the modified sun density $N_{e}\rightarrow \cos^2\theta_{13} N_e $. 
So the analyses of solar data constrain
three of the seven parameters: 
$\Delta m^2_{21}, \theta_{12}$ and $\theta_{13}$. The effect of $\theta_{13}$ is
to decrease the energy dependence of the solar survival probability.
\\
-- For atmospheric neutrinos, 
the solar wavelength is too long
and the corresponding oscillating phase is negligible. As a consequence
the atmospheric
data analysis restricts $\Delta m^2_{31}\simeq \Delta m^2_{32}$, 
$\theta_{23}$ and
$\theta_{13}$, the latter being the only parameter common to both solar 
and atmospheric neutrino oscillations and
which may potentially allow for some mutual influence. The effect of
$\theta_{13}$ is to add a $\nu_\mu\rightarrow\nu_e$ contribution to the
atmospheric oscillations. \\
-- At reactor experiments the solar wavelength is unobservable if 
$\Delta m^2< 8\times 10^{-4}$ eV$^2$ and the relevant survival probability
oscillates with wavelength determined by $\Delta m^2_{31}$ and  
amplitude determined by $\theta_{13}$. 

CP is unobservable in this approximation. There is, in principle some
dependence on the Normal versus Inverted orderings due to matter effects
in the Earth for atmospheric neutrinos, controlled by the mixing angle
$\theta_{13}$.  
In Fig.~\ref{teta13} I plot the results of the analysis 
of solar, atmospheric and reactor data on the allowed values of $\theta_{13}$.
The figure illustrates, that, at present, all data independently 
favours  $\theta_{13}=0$: the solar data exhibit energy
 dependence, the atmospheric data give no evidence for $\nu_e$
 oscillation, and, most important, reactor data exclude
 $\bar\nu_e$-disappearnce at the atmospheric wavelength
The combined analysis results in a limit
$\sin^2\theta_{13}\leq 0.06$ at 3$\sigma$~\cite{michele,review}. 
Within this limit the difference between Normal and Inverted orderings in 
atmospheric neutrino data is  below present experimental sensitivity.  
\begin{figure}
\begin{center}
\includegraphics[scale=0.35]{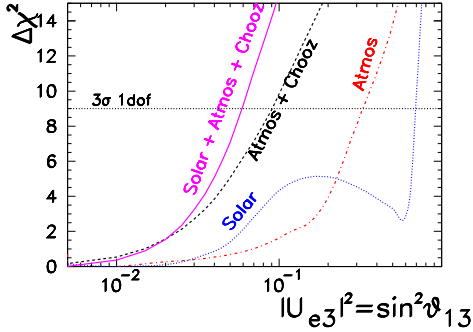}
\end{center}
\vglue -1cm
\caption{Dependence of $\Delta\chi^2$ on $\sin^2\theta_{13}$ in the analysis
of the atmospheric, solar and CHOOZ neutrino data. The dotted horizontal line 
corresponds to the 3$\sigma$ limit for a single parameter.}
\label{teta13}
\end{figure}

These results can be translated into our present knowledge of the
moduli of the mixing matrix $U$:
\begin{equation}
|U| =\pmatrix{ 0.73-0.89&0.45-0.66&<0.24 \cr
0.23-0.66&0.24-0.75&0.52-0.87\cr 
0.06-0.57&0.40-0.82&0.48-0.85\cr}
\end{equation}
which presents a structure 
\vskip -1cm
\begin{displaymath}
|U| \simeq \pmatrix{
\frac{1}{\sqrt{2}}(1+{ {\cal O}(\lambda)}) &
\frac{1}{\sqrt{2}}(1-{ {\cal O}(\lambda)}) &
{ \epsilon} \cr
-\frac{1}{{2}}(1-{ {\cal O}(\lambda)}+{{\epsilon}}) &
\frac{1}{{2}}(1+{ {\cal O}(\lambda)}-{{\epsilon}}) &
\frac{1}{\sqrt{2}}\cr
\frac{1}{{2}}(1-{ {\cal O}(\lambda)}-{{\epsilon}}) &
-\frac{1}{{2}}(1+{ {\cal O}(\lambda)}-{{\epsilon}}) &
\frac{1}{\sqrt{2}}\cr}
\end{displaymath}
with ${\lambda\sim 0.2}$ and ${\epsilon<0.25}$.
This structure is very different from that of the quark sector
\begin{displaymath}
|{ U_{\rm CKM}}| \simeq \pmatrix{1 & { {\cal O}(\lambda)}
& { {\cal O}(\lambda^3)} \cr
 { {\cal O}(\lambda)} & 1 &  { {\cal O}(\lambda^2)}\cr
 { {\cal O}(\lambda^3)} & { {\cal O}(\lambda^2)} & 1\cr}
\end{displaymath}
A good fraction of the papers of the last two years in Fig.~\ref{papers} is
devoted to find theoretical viable models {\sl explaining} (or more accurately
{\sl accommodating}) this fact.
\subsection{LSND and Sterile Neutrinos}
Together with the results from the solar and atmospheric neutrino  
experiments we have one more piece of 
evidence pointing out toward the existence of  
neutrino masses and mixing: the LSND  
results which finds evidence of $\overline\nu_\mu\rightarrow\overline\nu_e$ 
with $\Delta m^2\geq 0.1$ eV$^2$ (see Fig.~\ref{fig:miniboone}).
All data can be accommodated 
in a single neutrino oscillation framework only if there are at least 
three different scales of neutrino mass-squared differences which
requires the existence of a fourth light neutrino.
The measurement of the decay width of the
$Z^0$ boson into neutrinos makes the existence of three, and only three, light 
active neutrinos an experimental fact. Therefore, the fourth neutrino 
must not couple to the standard electroweak current, that is, it must be sterile.

One of the most important issues in the context of four-neutrino scenarios is 
the four-neutrino mass spectrum. There are six possible four-neutrino schemes,
shown in Fig.~\ref{fig:4mass}, that can accommodate the results from solar and 
atmospheric neutrino experiments as well as the LSND result. They can be 
divided in two classes: (3+1) and (2+2).
In the (3+1) schemes, there is a group of three close-by neutrino masses that 
is separated from the fourth one by a gap of the order of 1~eV$^2$, which is 
responsible for the SBL oscillations observed in the LSND experiment. 
In (2+2) schemes, there are two pairs of close masses separated by the LSND 
gap. The main difference between these two classes is the following: if a
(2+2)-spectrum is realized in nature, the transition into the sterile neutrino
is a solution of either the solar or the atmospheric neutrino problem, 
or the sterile neutrino takes part in both, whereas with a (3+1)-spectrum the
sterile neutrino could be only slightly mixed with the active
ones and mainly provide a description of the LSND result.
\begin{figure}
\includegraphics[scale=0.55]{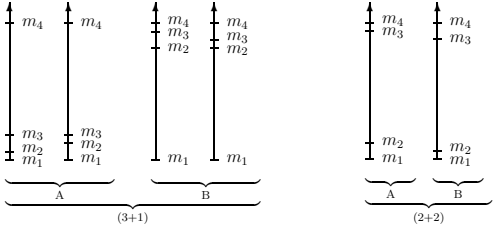}
\vglue -1cm
\caption{The six types of 4-neutrino mass spectra.} 
\label{fig:4mass}
\end{figure}

The phenomenological situation at present is that none of the four-neutrino
scenarios are favoured by the data as it was reviewed in the talk by 
Carlo Giunti~\cite{carlo} in the neutrino session. In brief 
(3+1)-spectra are disfavoured by the incompatibility between the LSND
signal and the present constraints
from short baseline laboratory experiments, while (2+2)-spectra are disfavoured by
the existing constraints from the sterile oscillations in solar and
atmospheric data. 
\begin{figure}[ht]
\begin{center}
\mbox{\epsfig{file=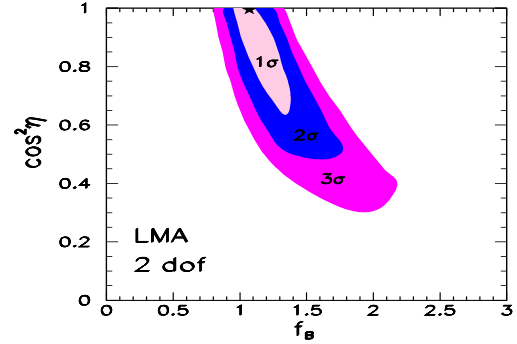,width=0.45\textwidth,height=0.3 \textwidth} }
\mbox{\epsfig{file=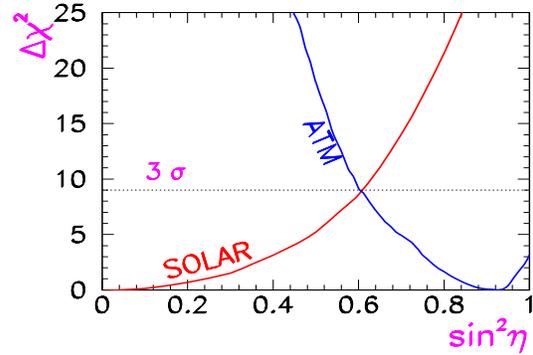,width=0.45\textwidth,height=0.3 \textwidth} }
\end{center}
\vglue -1cm
\caption{{\sl Upper}: Constraint on the active-sterile admixture in solar neutrino
oscillations versus the $^8$B neutrino flux enhancement factor. 
{\sl Lower}: Present status of the bounds on the active-sterile admixture from
solar (from Ref.~\cite{sterile}) and atmospheric (from Ref.~\cite{valle}) 
neutrino data in (2+2)-models.}
\label{ster}
\end{figure}

In this respect it has been recently pointed out that
the existing constraint on the sterile admixture in the solar neutrino
oscillations can be relaxed if the $^8$B neutrino flux is allowed to be
larger than in the SSM by a factor $f_B$~\cite{barger,sterile}. 
The analysis  is performed in
the context of  solar conversion  $\nu_e
\rightarrow \nu_x$, where $\nu_x = \cos \eta~ \nu_a + \sin \eta ~\nu_s$.
In Fig.~\ref{ster} I show the presently allowed range of $\eta$ 
as a function of $f_B$. The obtained upper bound on $\sin^2\eta$ from this
most general solar analysis has to be compared with the corresponding lower
bound from the analysis of atmospheric data. In Fig.~\ref{ster} I show the
corresponding comparison (the curve for the atmospheric data is taken from 
Ref.~\cite{valle}).

Alternative explanations to the LSND observation include the 
possibility of
CPT violation~\cite{CPT} which would imply that the mass differences and
mixing  among neutrinos would be different from the ones for antineutrinos,
or the possibility of lepton number violation in $\mu$ decay: 
$\mu^+\rightarrow e^+ \overline{\nu_e}\overline{\nu_i}$~\cite{babu}.  
Again, we find ourselves  in the privileged situation of having
experiments running which will be able to test these scenarios. 
For example an imminent test of CPT will be the comparison of 
the observation in KamLAND of $\overline{\nu_e}$ disappearance versus 
solar $\nu_e$ disappearance. Also, at present, MiniBooNE~\cite{miniboone}  
is running in the neutrino mode  searching for $\nu_\mu\rightarrow\nu_e$ to
be compared with the antineutrino signal in LSND. Thus  
an oscillation  signal at KamLAND or MiniBooNE will
put serious constraints on CPT violation for $\nu$'s. 
Further precision tests can be
performed at future facilities such as $\nu$ factories~\cite{ohlsson}.
\section{Implications}
\label{impli}
\subsection{The Need of New Physics}
The SM is based on the gauge symmetry
$SU(3)_{\rm C}\times SU(2)_{\rm L}\times U(1)_{\rm Y}$  
spontaneously broken to $SU(3)_{\rm C}\times U(1)_{\rm EM}$ by the 
the vacuum expectations value (VEV), $v$, of the a Higgs doublet 
field $\phi$.
The SM contains three fermion generations which reside in chiral 
representations of the gauge group. Right-handed fields are included 
for charged fermions as they are needed to build the electromagnetic
and strong currents.  No right-handed neutrino is included in the
model since neutrinos are neutral. 

In the SM, fermion masses arise from the Yukawa interactions,
\begin{eqnarray}
-{\cal L}_{\rm Yukawa}&=&Y^d_{ij}\overline{Q_{Li}}\phi D_{Rj}+
Y^u_{ij}\overline{Q_{Li}}\tilde\phi U_{Rj}+ \nonumber \\
&& Y^\ell_{ij}\overline{L_{Li}}\phi E_{Rj}+{\rm h.c.},
\label{yukawa}  
\end{eqnarray}
(where $\tilde\phi=i\tau_2\phi^\star$) after spontaneous symmetry breaking. The
Yukawa interactions of Eq.~(\ref{yukawa}) lead to charged fermion masses 
\begin{equation}
m^f_{ij}=Y^f_{ij} \frac{v}{\sqrt{2}}
\label{diracmass}
\end{equation}
but leave the neutrinos massless. No Yukawa interaction can be written that
would give mass to the neutrino because no right-handed neutrino field
exists in the model. 

One could think that neutrino masses would arise 
from loop corrections if these corrections induced effective terms
\begin{equation}
{Y^\nu_{ij}\over v} 
\left(\bar L_{Li} \tilde \phi\right)
\left( {\tilde\phi}^T L^C_{Lj}\right) +{\rm h.c.},
\label{radcor}  
\end{equation}
($L^C_{Lj}=C {\bar L_{Lj}}^T$).
This, however, cannot happen, as can be easily understood by examining the
accidental symmetries of the Standard Model. 
Within the SM the following accidental global symmetry arises:
\begin{equation}
G_{\rm SM}^{\rm global}=U(1)_B\times U(1)_e\times U(1)_\mu\times U(1)_\tau.
\label{SMglob}  
\end{equation}
Here $U(1)_B$ is the baryon number symmetry, and $U(1)_{e,\mu,\tau}$
are the three lepton flavor symmetries, with total lepton number given by
$L=L_e+L_\mu+L_\tau$.
Terms of the form (\ref{radcor}) violate $G_{\rm SM}^{\rm global}$ and 
therefore cannot be induced by loop corrections. Furthermore, the
$U(1)_{B-L}$ subgroup of $G_{\rm SM}^{\rm global}$ is non-anomalous.
Terms of the form (\ref{radcor}) have $B-L=-2$ and therefore cannot be
induced even by non-perturbative corrections.

It follows that the SM predicts that neutrinos are precisely
massless. Consequently, there is neither mixing nor CP violation in the
leptonic sector. Thus the simplest and most straightforward 
lesson of the evidence
for neutrino masses is also the most striking one: {\sl there is
new physics beyond the SM}. This is the first
experimental result that is inconsistent with the SM.
\subsection{The Scale of New Physics} 
There are many good reasons to think that the SM is not a
complete picture of Nature and some new physics (NP) is expected to
appear at higher energies. In this case the SM is an effective 
low energy theory valid  up to the scale $\Lambda_{\rm NP}$ 
which characterizes the NP. In this approach, 
the gauge group, the fermionic spectrum, and the pattern of 
spontaneous symmetry breaking 
are still valid ingredients to describe Nature at 
energies $E\ll\Lambda_{\rm NP}$. 
The difference between the SM as a complete description of Nature
and as a low energy effective theory is that in the latter case we must
consider also non-renormalizable (dim$>4$) terms whose effect 
will be suppressed by powers $1/\Lambda_{\rm NP}^{\rm dim-4}$. 
In this approach the largest effects at low energy are expected 
to come from dim$=5$ operators 

There is a single set of dimension-five terms that is made of 
SM fields and is consistent with the gauge symmetry  given by
\begin{equation}
{\cal O}_5={Z^\nu_{ij}\over 2 \Lambda_{\rm NP}}
\left(\bar L_{Li} \tilde \phi\right)
\left( {\tilde \phi}^T L^C_{Lj}\right)+{\rm h.c.},
\label{dimfiv}  
\end{equation}
which violates  total lepton number by two units and leads, upon 
spontaneous symmetry breaking, to neutrino masses:
\begin{equation}
(M_\nu)_{ij}={Z^\nu_{ij}\over 2}{v^2\over\Lambda_{\rm NP}}.
\label{nrmass}  
\end{equation}
This is a Majorana mass term. 

Eq.~(\ref{nrmass}) arises in a generic extension of the
SM which means that neutrino masses are very likely to appear
if there is NP. Furthermore comparing Eq.~(\ref{nrmass}) and
Eq.~(\ref{diracmass}) we find that 
the scale of neutrino masses 
is suppressed by $v/\Lambda_{\rm NP}$  when compared to the scale 
of charged fermion masses providing an explanation not only for 
the existence of neutrino masses but also for their smallness. 
Finally, Eq.~(\ref{nrmass}) breaks not only total lepton number but also
the lepton flavor symmetry $U(1)_e\times U(1)_\mu\times U(1)_\tau$.
Therefore we should expect lepton mixing and CP violation.

Given the relation (\ref{nrmass}), $m_\nu\sim v^2/\Lambda_{\rm NP}$, it is
straightforward to use measured neutrino masses to estimate the scale
of NP that is relevant to their generation. In particular,
if there is no quasi-degeneracy in the neutrino masses, the heaviest
of the active neutrino masses can be estimated,
\begin{equation}
m_h=m_3\sim\sqrt{\Delta m^2_{\rm atm}}\approx 0.05\ {\rm eV}.
\label{estmth}
\end{equation}
(In the case of inverted  hierarchy the implied scale is 
$m_h=m_2\sim\sqrt{|\Delta m^2_{\rm atm}|}\approx 0.05\ {\rm eV}$). 
It follows that the scale in the non-renormalizable term (\ref{dimfiv})
is given by
\begin{equation}
\Lambda_{\rm NP}\sim v^2/m_h\approx10^{15}\ {\rm GeV}.
\label{estlnp}
\end{equation}
We should clarify two points regarding Eq.~(\ref{estlnp}):

1. There could be some level of degeneracy between the neutrino masses that 
are relevant to the atmospheric neutrino oscillations. In such a case 
Eq.~(\ref{estmth}) is modified into a lower bound
and, consequently, Eq.~(\ref{estlnp}) becomes an upper bound on
the scale of NP.

2. It could be that the $Z_{ij}$ couplings of Eq.~(\ref{dimfiv}) are much
smaller than one. In such a case, again, Eq.~(\ref{estlnp}) becomes an upper 
bound on the scale of NP. On the other hand, in models of approximate flavor 
symmetries,
there are relations between the structures of the charged lepton and neutrino
mass matrices that give quite generically  $Z_{33}\geq m_\tau^2/v^2\sim
10^{-4}$. We conclude that the likely range of $\Lambda_{\rm NP}$
that is implied by the atmospheric neutrino results is given by
\begin{equation}
10^{11}\ {\rm GeV}\leq\Lambda_{\rm NP}\leq 10^{15}\ {\rm GeV}.
\label{ranlnp}
\end{equation}

The estimates (\ref{estlnp}) and (\ref{ranlnp}) are very exciting.
First, the upper bound on the scale of NP is well below the
Planck scale. This means that there is a new scale in Nature which
is intermediate between the two known scales, the Planck scale
$m_{\rm Pl}\sim10^{19}$ GeV and the electroweak breaking scale,
$v\sim10^2$ GeV. 

It is amusing to note in this regard that the solar neutrino problem does not 
necessarily imply such a new scale. If its solution is related to vacuum 
oscillations with $\Delta m^2_{21}\sim10^{-10}$ eV$^2$, it can be explained 
by $\Lambda_{\rm NP}\sim m_{\rm Pl}$. 
However, the 
favoured explanation for the solar neutrino deficit is the LMA solution which 
again points towards NP scale in the range of Eq.~(\ref{ranlnp}).

Second, the scale $\Lambda_{\rm NP}\sim10^{15}$ GeV is intriguingly close
to the scale of gauge coupling unification. 

Of course, neutrinos could be conventional Dirac particles.
In the minimal realization of this possibility, one must still extend
the SM to add right-handed neutrinos and {\sl impose} the conservation 
of total lepton number (since in the presence of right-handed neutrinos
total lepton number is not an accidental symmetry) to prevent
the right-handed neutrinos from acquiring a singlet Majorana mass term.
In this scenario, neutrinos could acquire a mass like any other fermion
of the Standard Model and no NP scale would be implied. 
We would be left in the darkness on the reason of the smallness 
of the neutrino mass.  

\subsection{Reconstructing the Neutrino Mass Matrix}
The best known scenario that leads to (\ref{dimfiv}) is the {\it see-saw
mechanism}~\cite{seesaw}. 
Here one
assumes the existence of heavy sterile neutrinos $N_i$. Such fermions
have SM gauge invariant bare mass terms and Yukawa interactions :
\begin{equation}
-{\cal L}_{\rm NP}=\frac{1}{2} {M_N}_{ij}\overline{N^c_i}N_j+
Y^\nu_{ij}\overline{L_{Li}}\tilde\phi N_j +{\rm h.c.}.
\label{sinint}  
\end{equation}
The resulting mass matrix 
in the basis $\left(\begin{array}{c}\nu_{Li}\\ N_j\end{array}\right)$  
has the following form:
\begin{equation}
M_\nu=\pmatrix{0&Y^\nu{v\over\sqrt2}\cr (Y^\nu)^T{v\over\sqrt2}&M_N\cr}.
\label{fumama}  
\end{equation}
If the eigenvalues of $M_N$ are all well above the electroweak breaking scale
$v$, then the diagonalization of $M_\nu$ leads to three light mass eigenstates
and an effective low energy interaction  of the form (\ref{dimfiv}). 
In particular, the scale
$\Lambda_{\rm NP}$ is identified with the mass scale of the heavy sterile
neutrinos, that is the typical scale of the eigenvalues of $M_N$.
Two well-known examples of extensions of the SM that lead to
a see-saw mechanism for neutrino masses are SO(10) GUTs
and left-right symmetry.

One may notice that even in this particularly simple form of NP,  ${\cal L}_{\rm NP}$ 
contains  18 parameters which we would need to know in order to fully
determine the dynamics of the NP. However the effective low energy operator
${\cal O}_5$ contains only 9 parameters which we can hope to measure at
the low energy experiments.  
This  simple parameter counting illustrates the limitation of the 
``bottom-up'' approach in deriving model independent implications of the
presently observed neutrino masses and mixing. 

Alternatively one can go  ``top-down'' by studying the low energy 
effective neutrino masses and mixing induced by specific high
energy models~\cite{review}. For example, in his talk, Q. Shafi~\cite{shafi}
discussed the possibility of accommodating the present neutrino data
in the framework of theories with warped extra dimensions.
Unfortunately the number of possible models is 
overwhelming and impossible to review in this talk. 

An intermediate approach is the attempt to classify
different forms of NP in terms of the characteristic 
texture of the induced low energy neutrino mass matrices~\cite{altarelli}. 
The main goal is to identify generic
predictions which, with more data at hand, will be able to discriminate
among the different textures. In general, depending on the specific
texture, different relations between the mass differences and the
mixing angles are expected. For example relations between 
the value of $\theta_{13}$ and the ratio of the masses, and
different rate for  neutrinoless double beta decay,   
which could be tested once the parameters are known to good 
enough precision. 

The bottom line is that in order to advance further in this
direction we need more  (and more precise) data. As we will
see  even at the end of the program of the presently
approved experiments we will still be far from reaching this
goal.

\subsection{A {\sl Side Effect:} Leptogenesis}
Finally I would like to comment a possible {\sl side effect}
\footnote{I call it side effect because 
it is not guaranteed to happen.}  
of neutrino masses which is that they may help us to explain 
``how we are here''. What I mean with this, is the explanation of  
the origin of the cosmic matter-antimatter asymmetry
via leptogenesis~\cite{lepto}.   
From the Big-Bang Nucleosynthesis, we know
that there is only a tiny asymmetry in the baryon number,
$n_B/n_\gamma \approx 5 \times 10^{-10}$.  
Leptogenesis~\cite{lepto} is the possible origin of
such a small asymmetry related to  neutrino physics.
\begin{figure}
\begin{center}
\includegraphics[width=1.1\columnwidth]{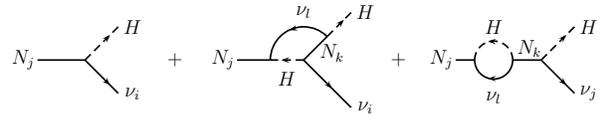}
\vglue -1.cm
\caption{The tree-level and one-loop diagrams of
right-handed neutrino decay into leptons and Higgs.}  
\label{fig:lepto}
\end{center}
\end{figure}

In a possible realization of leptogenesis, 
$L \neq 0$ is generated  in  the 
Early Universe by the decay of one of the heavy  right-handed neutrinos
of the see-saw mechanism, with a direct CP
violation. Due to the
interference between the tree-level and one-loop diagrams shown 
in Fig.~\ref{fig:lepto}
the decay rates of the right-handed neutrino into leptons
and anti-leptons are different. In order to generate a
lepton asymmetry the decay must be out of equilibrium 
($\Gamma_{\nu_R}\ll$ Universe expansion rate).
Sphaleron processes transform the lepton asymmetry into a  
baryon asymmetry and below the electroweak phase transition 
a net baryon asymmetry is generated
$\Delta B\simeq -\frac{\Delta \rm L}{2}$ 
(the exact coefficient relating $\Delta B $
to ${\Delta \rm L}$ is model dependent.)

The details of the leptogenesis scenario are model dependent
and much work has been done in the framework of specific neutrino
models (see, for example, the talk by Xing [39]). In particular,
it has been shown that a right-handed neutrino 
of about $10^{10}$~GeV  can account for the cosmic baryon asymmetry from its 
out-of-equilibrium decay~\cite{leptoreview}. 

\section{Future}
\label{future}
\subsection{KamLAND and Borexino}
Our present understanding of the solar neutrino oscillation is 
being tested in the 
KamLAND experiment  which is currently in operation in the 
Kamioka mine in Japan. This
underground site is conveniently located at a distance of 150-210 km 
from several Japanese nuclear power stations. The measurement of the flux and 
energy spectrum of the $\bar\nu_e$'s emitted by these reactors will 
provide a test to the LMA solution of the solar neutrino anomaly~\cite{kland}.
In Fig.~\ref{kamdist} I show the expected distortion on the energy spectrum
in the presence of oscillations. 
\begin{figure}
\begin{center}
\includegraphics[scale=0.6]{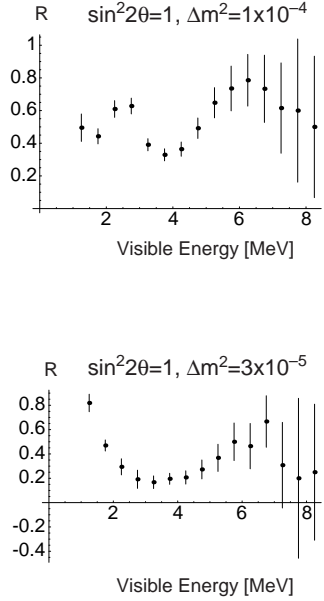}
\end{center}
\vglue -1cm
\caption{Predicted distortion in the energy spectrum of KamLAND in the
presence of oscillations, from Ref.~\cite{mura}.} 
\label{kamdist}
\end{figure}
After two or three  years of data taking, 
KamLAND should be capable of either excluding the entire LMA region or, 
not only establishing 
$\nu_{e}\leftrightarrow \nu_{\rm other}$ oscillations, but also 
measuring the LMA oscillation parameters with unprecedented 
precision~\cite{mura,kland2,sterile} provided that 
$\Delta m^2\leq {\rm few} 10^{-4}$ eV$^2$.
KamLAND is expected to announce their first results this year. 
\begin{figure}
\begin{center}
\includegraphics[scale=0.45]{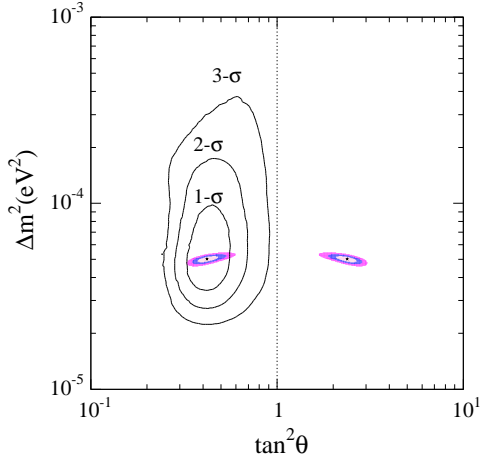}
\end{center}
\vglue -1cm
\caption{Expected reconstructed regions if KamLAND observes a signal
corresponding to the present best fit point of the LMA region.
From Ref.~\cite{sterile}.} 
\end{figure}

If LMA is confirmed, CP violation may be  observable at future 
long-baseline (LBL)
experiments. Also KamLAND will provide us, as data accumulate, with a firm 
determination of the corresponding oscillation parameters. 
With this at hand, the future solar $\nu$ experiments will be able to return
to their original goal of testing solar physics. 

If KamLAND does not confirm LMA, the next most relevant results will come from
Borexino ~\cite{borexino}. 
The Borexino experiment 
is designed to detect low-energy solar neutrinos in real-time through the 
observation of the ES process $\nu_a + e^- \to \nu_a + e^- $. The energy 
threshold for the recoil electrons is 250 keV. The largest contribution 
to their expected event rate is from neutrinos of the $^7$Be line.
Due to the lower energy threshold, Borexino is sensitive to matter effects in 
the Earth in the LOW region. And because $^7$Be neutrinos 
are almost monoenergetic, it is also very sensitive to seasonal 
variations associated with VAC oscillations. 

\subsection{Long Baseline Experiments}
$\nu_\mu$  oscillations with  $\Delta m^2_{\rm atm}$ are being probed
and will be further tested  using accelerator beams at 
LBL experiments. In these experiments the intense neutrino  
beam from an accelerator is aimed at a detector located underground at a 
distance of several hundred kilometers. 
At present there are three such projects approved: K2K~\cite{k2k} 
which runs with a baseline of about 
235 km from KEK to SK, MINOS~\cite{minos} 
under construction with a baseline of 730 km from Fermilab to 
the Soudan mine where the detector will be placed, 
and two detectors OPERA and ICARUS~\cite{opera,icarus}  
under construction with a baseline of 730 km from CERN to Gran Sasso.
\begin{figure}[ht]
\vglue -1cm
\begin{center}
\includegraphics[scale=0.45]{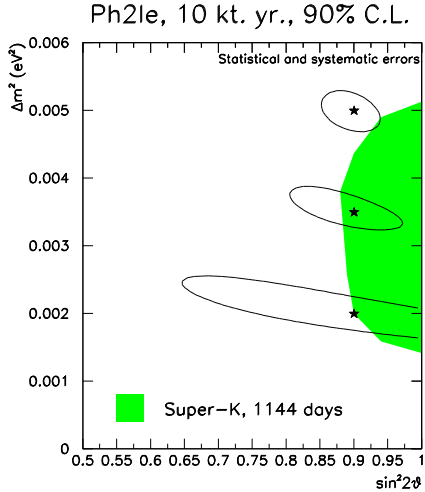}
\end{center}
\vglue -1cm
\caption{Expected reconstruction of the oscillation parameters in 
Minos. See Ref.~\cite{minos} for details.}
\label{fig:minos}
\end{figure}

The first results from K2K seem to confirm the atmospheric 
oscillations but statistically they 
are still not svery significant. In the near future K2K 
will accumulate more data enabling it to confirm the
 atmospheric neutrino oscillation. Furthermore, combining the
 K2K and atmospheric neutrino data will lead to 
a better determination of the mass and mixing parameters.

In a longer time scale, the results from MINOS will provide more 
accurate determination of these parameters as shown in Fig.~\ref{fig:minos}. 
OPERA and ICARUS are designed to observe the $\nu_\tau$ appearance. 
MINOS, OPERA and ICARUS have certain sensitivity to $\theta_{13}$ although
by how much they will be ultimately able to improve the present 
bound is still undetermined.
\subsection{MiniBOONE}
The MiniBooNE experiment~\cite{miniboone}, will be able to confirm 
or disprove the LSND oscillation signal within the next two years 
(see Fig.~\ref{fig:miniboone}). 
\begin{figure}[ht]
\vglue -0.5cm
\begin{center}
\includegraphics[scale=0.4]{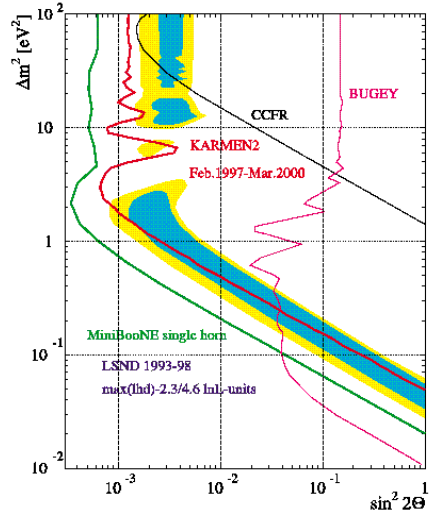}
\end{center}
\vglue -1cm
\caption{
Allowed regions (at 90 and 99 \% CL) for $\nu_e\to\nu_\mu$ 
oscillations from the LSND experiment compared with the exclusion regions 
(at 90\% CL) from KARMEN2 and other experiments. The 90 \% CL expected 
sensitivity curve for MinimBoNE is also shown.}
\label{fig:miniboone}
\end{figure}
Should the signal be confirmed as well as the solar signal in KamLAND and 
the atmospheric in LBL experiments, we will face the challenging 
situation of not having a successful  ``minimal'' phenomenological 
description at low-energy of the leptonic mixing. 

\subsection{Neutrino Mass Scale}
\label{direcdet}
Oscillation experiments provide information on
$\Delta m^2_{ij}$, 
and on the leptonic mixing angles, $U_{ij}$. But they are insensitive to 
the absolute mass scale for the neutrinos. 
Of course, the results of an oscillation experiment do provide a lower bound
on the heavier mass in $\Delta m^2_{ij}$, $|m_i|\geq\sqrt{\Delta m^2_{ij}}$ for
$\Delta m^2_{ij}>0$. But there is no upper bound on this mass. In particular,
the corresponding neutrinos could be approximately degenerate at a mass
scale that is much higher than $\sqrt{\Delta m^2_{ij}}$. 
Moreover, there
is neither upper nor lower bound on the lighter mass $m_j$.

Information on the neutrino masses, rather than mass differences, can be 
extracted from kinematic studies of reactions in which a neutrino or an 
anti-neutrino is involved. In the presence of mixing the most 
relevant constraint comes from Tritium beta decay 
${\rm ^3H \rightarrow\ ^3He + e^-+\overline\nu_e}$ 
which, within the present and expected
experimental accuracy, can limit the combination 
\begin{equation}
m_{\beta}=\sum_i m_i |U_{ei}|^2
\label{mbeta}
\end{equation}
The present bound is 
$m_{\beta}\leq 2.2$ eV at 95 \% CL~\cite{tritium}.
A new experimental project, KATRIN, is under consideration with an estimated 
sensitivity limit: $m_{\beta}\sim0.3$ eV~\cite{wark}. 

Direct information on neutrino masses can also 
be obtained from neutrinoless double beta decay
$(A,Z) \rightarrow (A,Z+2) + e^{-} + e^{-}$.
The rate of this process is proportional to the 
{\it effective Majorana mass of $\nu_e$},
\begin{equation}
m_{ee}=\left| \ \ \sum_i m_i U_{ei}^2 \ \ \right|
\end{equation}
which, unlike Eq.~(\ref{mbeta}),
depends also on the three CP violating phases. 
Notice that in order to induce the $2\beta0\nu$ decay, $\nu$'s must 
Majorana particles. 

The present strongest bound from $2\beta0\nu$-decay is 
$ m_{ee} < 0.34$ eV at 90 \% CL~\cite{klapdor}. 
Taking into account systematic errors related to nuclear matrix elements,
the bound may be weaker by a factor of about 3. A sensitivity of 
$m_{ee}\sim0.1$ eV is expected to be reached by the currently running 
NEMO3 experiment, while a series of new 
experiments (CUORE, EXO, GENIUS) is planned with sensitivity of up to  
$m_{ee} \sim 0.01$ eV~\cite{wark}. 

The knowledge of $m_{ee}$ can provide information on the mass
and mixing parameters that is independent of the $\Delta m^2_{ij}$'s. However,
to infer the values of neutrino masses, additional assumptions are required.
In particular, the mixing elements are complex and may lead to strong 
cancellation, $m_{ee}\ll m_1$. Yet, the combination of results
from $2\beta0\nu$ decays and Tritium beta decay can test and,
in some cases, determine the mass parameters of given
schemes of neutrino masses~\cite{bb} provided that the nuclear matrix elements
are known to good enough precision. 
\subsection{Future Facilities}
At the end of the presently approved neutrino experiments, many
questions will still remain open. Even in the scenario in 
which MiniBooNE does not confirm the LSND signal, and we can live 
with oscillations among the three known neutrinos, we will still
be ignorant about: (i) the value of $\theta_{13}$, (ii) 
the sign($\Delta m^2_{13}$), and (iii)  the possibility of CP violation 
in the lepton sector.

To measure these parameters, the following is required of future 
experiments:\\
(i) To measure $\theta_{13}$: Very intense beam with low
background\\
(ii) To discriminate Normal/Inverted:  
 Matter effects which implies very long baseline. \\
(iii) To detect CP violation: LMA must be confirmed and 
$\theta_{13}$ should be not too small. One must have 
intense beams with  exchangeable initial state
($\nu$/$\bar\nu$).

New facilities and experiments are being proposed which can realize some
of all of these conditions. In particular, for future neutrino oscillation 
experiments two type of facilities are being proposed: conventional 
neutrino superbeams~\cite{superbeams,jhf}  
(conventional meaning from the decay of pions generated 
from a  proton beam dump) with a detector either on or off axis, 
and neutrino beams from muon decay in muon storage rings~\cite{nufact}. 
As an illustration of the possible reach of these facilities I 
show in Fig.~\ref{mauro} the 
required values of $\Delta m^2_{21}$ and
mixing angle $\theta_{13}$ which would allow to measure a maximal
CP violation phase at different type of neutrino beams 
(from the CERN working group on Super Beams~\cite{jj})
\begin{figure}[ht]
\vglue -0.5cm
\includegraphics[scale=0.37,angle=90]{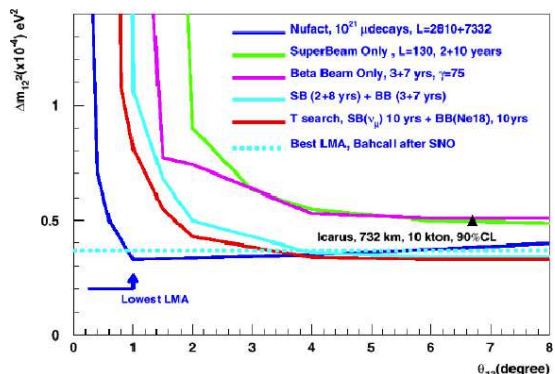}
\vglue -1cm
\caption{Required values of $\Delta m^2_{21}$ and
mixing angle $\theta_{13}$ which would allow to measure a maximal
CP violation phase at different types of neutrino beams and baselines.}
\label{mauro}
\end{figure}
See Ref.~\cite{wark} for a more exhaustive comparison of the expected 
reach of these  proposals.  

In general, the independent determination of these
missing pieces of the puzzle at these facilities becomes challenging
because  in the relevant oscillation probabilities 
there appear three independent two-fold parameter degeneracies
$(\delta_{CP}\,, \theta_{13})$, sign$(\Delta m^2_{31})$ and $(\theta_{23}\,,
\pi/2-\theta_{23})$. The phenomenological
efforts in this front concentrate on the study of how 
the combination of data from experiments performed at different baselines 
and different beam types can help in resolving these degeneracies
~\cite{degeneracies}.

\section{Conclusions}
Neutrino physics is a very exciting field which is at the moment 
experimentally driven. An enormous experimental effort has been 
devoted in the last years to prove beyond doubt the presence of
neutrino masses and mixing. In the last year, the SNO experiment 
has provided model independent evidence of flavour
conversion of solar $\nu_e's$ at more than 5 $\sigma$ CL. 
SuperKamiokande has accumulated more and more data on the
disappearance of atmospheric $\nu_\mu$'s resulting into a  
high confidence of the $L/E$ dependence of their survival probability.

At present these signals are being probed and will be further tested 
with ``human-made'' neutrino beams from reactor and accelerators. 
For solar $\nu's$  KamLAND should  give us a definite answer on the
LMA solution to the solar neutrino problem and provide us with 
a precise determination of the corresponding oscillation parameters.
The present K2K data seem to confirm the atmospheric neutrino 
oscillation within the limited statistics. The experiments should soon
recover from the accident in SuperKamiokande, and with additional
data will provide a more meaningful test of the oscillation.
In the near future MINOS will measure the oscillation parameters 
with precision. Within the next two years MiniBooNE will definitively 
test the LSND signal. 

Neutrino masses imply physics beyond the Standard Model and 
they most probably suggest a NP
scale close to GUT scale. The implied existence of heavy, 
SM-singlet neutrinos opens up the possibility of leptogenesis 
as the mechanism for generation of the baryon asymmetry.  

Determining the parameters of the neutrino mass matrix 
will provide fundamental information to understand 
the dynamics at the new physics scale.
However even at the end of the existing neutrino programs, 
we will still be far from reaching this goal.
Further advance requires a new generation of neutrino 
experiments.
\vspace*{-0.3cm}
\section*{Acknowledgements}
\vskip -0.1cm
I want to give special thanks to 
J.N. Bahcall, J.J. Gomez-Cadenas,  P. Hernandez, M. Maltoni, A. Marrone 
and C. Pe\~na-Garay, for their help in the preparation of my talk.
This work is supported by 
the MC fellowship HPMF-CT-2000-00516 and by DGICYT grant FPA2001-3031. 


\begin{thebibliography}{9}
\bibitem{wark} David Wark in these proceedings.
\bibitem{sno} Scott Oser in these proceedings.
\bibitem{sksolar} Mark R. Vagins in these proceedings.
\bibitem{kland} Tadao Mitsui in these proceedings.
\bibitem{skatm} Christopher Mauger in these proceedings.
\bibitem{k2k} Yoshinari Hayato in these proceedings.
\bibitem{jhf} Y. Itow in these proceedings.
\bibitem{miniboone} Andrew Bazarko in these proceedings
\bibitem{minos} J. Urheim in these proceedings.
%
\bibitem{MNS}
Z. Maki, M. Nakagawa, and S. Sakata, 
Prog. Theor. Phys. {\bf 28}, 870 (1962).
%
\bibitem{ckm}
Kobayashi, M., and T.~Maskawa,, 
Prog.\ Theor.\ Phys.\  {\bf 49}, 652 (1973).
%
\bibitem{pontecorvo}
B. Pontecorvo,  
J. Exptl. Theoret. Phys. {\bf 33}, 549 (1957). 
\bibitem{msw}
L. Wolfenstein, 
Phys. Rev. D {\bf 17}, 2369 (1978);
S.P. Mikheyev, and A.Y. Smirnov,  
Yad. Fiz. {\bf 42}, 1441 (1985) [Sov. J. Nucl. Phys. {\bf 42}, 913].
%
\bibitem{dark}
A. de Gouvea, A. Friedland and H. Murayama,
Phys.\ Lett.\ B {\bf 490}, 125 (2000);
 M. C. Gonzalez-Garcia, C. Pe\~na-Garay, 
Phys. Rev. {\bf D62}, 031301 (2000);
G. L. Fogli, E. Lisi, A. Maronne and G. Scioscia,
 Phys. Rev. {\bf D59}, 033001 (1999);
 C. Giunti, M. C. Gonzalez-Garcia, C. Pe\~na-Garay, 
Phys. Rev. {\bf D62}, 013005 (2000).
%
\bibitem{bp00}
J.N. Bahcall, H.M. Pinsonneault, and S. Basu, 
Astrophys.\ J.\  {\bf 555}, 990 (2001).
%
\bibitem{solana}
SNO coll., nucl-ex/0204009;
Barger {\sl et al.}, hep-ph/0204253; 
Bandyopadhysy {\sl et al},  hep-ph/0204286;
Bahcall  {\sl et al}, hep-ph/0204314;
Creminelli {\sl et al}, hep-ph/0102234;
Aliani {\sl et al} hep-ph/0205053;
de Holanda, Smirnov,  hep-ph/0205241;
Strumia {\sl et al},  hep-ph/0205262;
Fogli {\sl et al}, hep-ph/0206162;
Maltoni {\sl et al},hep-ph/0207227.
%
\bibitem{oursolar}
J.N. Bahcall, M.C. Gonzalez-Garcia, C.
Pe\~na-Garay,  JHEP {\bf 0207}, 054 (2002).
%
\bibitem{ichep00}
M.C. Gonzalez-Garcia,  Osaka 2000, High energy
physics, vol. 2 899-906, hep-ph/0010136 
%
\bibitem{solarprobs}
V. S. Berezinsky, M. Lissia, hep-ph/0108108;
G. L. Fogli et al., hep-ph/0203138;
V. Barger, D. Marfatia, K. Whisnant, B. Wood,
hep-ph/0204253. 
%
\bibitem{sfp}
C.S. Lim,  W. J. Marciano, Phys. Rev. D37 1369 (1988); 
E. Kh Akhmedov, Phys. Lett. B213, 64 (1988).
%
\bibitem{sfpana}
E. K. Akhmedov, J. Pulido,  
Phys. Lett. {\bf B529}, 193 (2002);
A. M. Gago, et al, Phys. Rev. {\bf D65} 073012 (2002);
O. G. Miranda et al., hep-ph/0108145;
A. Friedland, A. Gruzinov, hep-ph/0202095;   
B. C. Chauhan, J. Pulido hep-ph/0206193;   
J. Barranco, O.G. Miranda, T.I. Rashba, V.B. Semikoz, 
J.W.F. Valle, hep-ph/0207326. 
%
\bibitem{fcnc}
L. Wolfenstein, Phys. Rev. {\bf D17} 2369 (1978)
E.  Roulet, Phys. Rev. {\bf D44} R935 (1991); 
%
\bibitem{fcncana}
M. M. Guzzo, A. Masiero, S. T. Petcov, Phys. Lett. {\bf B260},154 (1991);
Guzzo {\sl et al} hep-ph/0112310; 
Gago {\sl et al} hep-ph/0112060.
%
\bibitem{michele} M.~C.~Gonzalez-Garcia and M.~Maltoni,
hep-ph/0202218.
%
\bibitem{fogli} Fogli, Lisi and Marrone,
Phys.\ Rev.\ D {\bf 64}, 093005 (2001).

\bibitem{chooz} CHOOZ Collaboration, M. Apollonio {\it et al.},
Phys.Lett. {\bf B420}, 397 (1998).
%
\bibitem{review}
M.C. Gonzalez-Garcia and Y. Nir, hep-ph/0202058, Rev. Mod. Phys, in press.
%
\bibitem{carlo} C. Giunti in these proceedings.
%
\bibitem{barger}
V. Barger, D. Marfatia, and K. Whisnant,
Phys. Rev. Lett. {\bf 88}, 011302 (2002)
%
\bibitem{sterile}
J.~N.~Bahcall, M.~C.~Gonzalez-Garcia and C.~Pena-Garay,
Phys.\ Rev.\ C {\bf 66}, 035802 (2002).
%
\bibitem{valle}
Maltoni, M., M.A. Tortola, T.~Schwetz and J.W.~Valle, 2002,
hep-ph/0207227. 
%
\bibitem{CPT}
H. Murayama, and T.~Yanagida, ,
Phys.\ Lett.\ B {\bf 520}, 263 (2001);
S. Pakvasa, hep-ph/0110175;
Barenboim {\em et al.},  hep-ph/0108199;   
Skadghauge,  hep-ph/0112189; 
A. Strumia, hep-ph/0201134.
%
\bibitem{babu} K.S. Babu in these proceedings.
%
\bibitem{ohlsson} T. Ohlsson in these proceedings.
 
\bibitem{seesaw}
P. Ramond, 
CALT-68-709 (1979);  
M.P. Gell-Mann, P. Ramond and R. Slansky, 1979, in {\it Supergravity}, 
edited by P. van Nieuwenhuizen and D.Z. Freedman (North Holland);
T. Yanagida, 1979, in {\it Proceedings of Workshop on Unified Theory and 
Baryon Number in the Universe}, edited by O. Sawada and A. Sugamoto (KEK);
R.N. Mohapatra, and G. Senjanovic, 1980, 
Phys. Rev. Lett. {\bf 44}, 912.

\bibitem{shafi} Q. Shafi in these proceedings.

\bibitem{altarelli}
For a recent review see 
G. Altarelli and F. Feruglio, hep-ph/0206077, and references therein. 
%
\bibitem{lepto}
M.~Fukugita and T.~Yanagida,
Phys.\ Lett.\ B {\bf 174}, 45 (1986).
%
\bibitem{leptoreview}
See for example,  W.~Buchmuller, hep-ph/0107153. 
talk at Presented at 8th International Symposium on 
Particle Strings and Cosmology (PASCOS 2001), Chapel Hill, North
Carolina, 10-15, 2001. 
%
\bibitem{xing} See Z. Xing in these proceedings.
%
\bibitem{kland2} 
V.~D.~Barger, D.~Marfatia and B.~P.~Wood,
Phys.\ Lett.\ B {\bf 498}, 53 (2001);
A.~de Gouv\^ea and C.~Pe\~na-Garay,
Phys.\ Rev.\ D {\bf 64}, 113011 (2001)
%
\bibitem{mura}
H.~Murayama and A.~Pierce,
Phys.\ Rev.\ D {\bf 65}, 013012 (2002)
%
\bibitem{borexino}
L. Oberauer, 
Nucl. Phys. Proc. Suppl. {\bf 77}, 48 (1999).
%
\bibitem{opera}
A.G. Cocco, {\em et al.},  OPERA Coll.,
Nucl.\ Phys.\ Proc.\ Suppl.\  {\bf 85}, 125 (2000).
%
\bibitem{icarus}
O.~Palamara, {\em et al.},  ICARUS Coll,,
Nucl.\ Phys.\ Proc.\ Suppl.\  {\bf 110}, 329 (2002).
%

\bibitem{tritium}
J. Bonn, {\em et al.},
Nucl.\, Phys.\, Proc.\, Suppl.\, {\bf 91}, 273 (2001);
V.M. Lobashev, {\em et al.}, 
Nucl.\, Phys.\, Proc.\, Suppl.\, {\bf 91}, 280 (2001).
%
\bibitem{klapdor}
H.V. Klapdor-Kleingrothaus, H.V.,
Eur.\ Phys.\ J.\ A {\bf 12} 147 (2001).
%
\bibitem{bb}
F. Vissani, JHEP {\bf 9906}, 022 (1999);
Y. Farzan, O.~L.~Peres and A.~Y.~Smirnov,
Nucl.\ Phys.\ B {\bf 612}, 59 (2001);
S.M. Bilenky, S.M., S.~Pascoli and S.~T.~Petcov,
Phys.\ Rev.\ D {\bf 64}, 053010 (2001);
{\sl ibid} 113003;S. Pascoli, and T.S. Petcov
hep-ph/0205022.

\bibitem{degeneracies}
J.~Burguet-Castell, M.~B.~Gavela, 
J.~J.~Gomez-Cadenas, P.~Hernandez and O.~Mena,
Nucl.\ Phys.\ B {\bf 608}, 301 (2001) and hep-ph/0207080;
V.~Barger, D.~Marfatia and K.~Whisnant,
Phys.\ Rev.\ D {\bf 65}, 073023 (2002);
P.~Huber, M.~Lindner and W.~Winter,
hep-ph/0204352.

\bibitem{superbeams}
B.~Richter, hep-ph/0008222;
V.~D.~Barger, S.~Geer, R.~Raja and K.~Whisnant,
Phys.\ Rev.\ D {\bf 63}, 113011 (2001);
J.~J.~Gomez-Cadenas {\it et al.}, hep-ph/0105297;
Y. Itow, {\it et al.}, hep-ex/0106019.
%
\bibitem{nufact}
S. Geer, Phys. ReV. {\bf D57} 6989 (1998).
A.~De Rujula, M.~B.~Gavela and P.~Hernandez,
Nucl.\ Phys.\ B {\bf 547}, 21 (1999);
J.~J.~Gomez-Cadenas and D.~A.~Harris,
FERMILAB-PUB-02-044-T.
%
\bibitem{jj} 
A. Blondel {\em et al.}, CERN-NUFACT-NOTE-095, 
CERN-OPEN-2002-025
\end{thebibliography}
\end{document}